\documentstyle[12pt,graphicx]{article}
\input colordvi
\def\Journal#1#2#3#4{{#1} {\bf #2}, #3 (#4)}
\def\NPB{{\em Nucl. Phys.} B}
\def\PLB{{\em Phys. Lett.}  B}
\def\PRL{\em Phys. Rev. Lett.}
\def\PRD{{\em Phys. Rev.} D}

\def\r2{\sqrt 2}
\def\beq{\begin{equation}}
\def\eeq{\end{equation}}
\def\beqn{\begin{eqnarray}}
\def\eeqn{\end{eqnarray}}

\def\sinW2{\sin^2\theta_W}

\def\mz2{M_{z}^2}
\def\c2b{\cos 2\beta}

\def\mz{M_z}

\def\Fq2{F_{2}(q^2)}
\def\beq{\begin{equation}}
\def\eeq{\end{equation}}

\def\gmin2{(g-2)_\mu}

\def\sec2w{sec^2\theta_W}
\def\r2{\sqrt 2}
\def\beq{\begin{equation}}
\def\eeq{\end{equation}}
\def\beqn{\begin{eqnarray}}
\def\eeqn{\end{eqnarray}}

\def\sinW2{\sin^2\theta_W}

\def\mz2{M_{z}^2}
\def\c2b{\cos 2\beta}

\def\mz{M_z}

\def\Fq2{F_{2}(q^2)}
\def\sq2{\sqrt{2}}

\def\sec2w{sec^2\theta_W}
\begin{document}

\begin{titlepage}

\begin{center}
{\large {\bf PHASES AND CP VIOLATION IN SUSY}}\\
\vskip 0.5 true cm
\vspace{2cm}
\renewcommand{\thefootnote}
{\fnsymbol{footnote}}
 Tarek Ibrahim$^{a,b}$ and Pran Nath$^{b}$  
\vskip 0.5 true cm
\end{center}

\noindent
{a. Department of  Physics, Faculty of Science,
University of Alexandria,}\\
{ Alexandria, Egypt\footnote{: Permanent address of T.I.}}\\ 
{b. Department of Physics, Northeastern University,
Boston, MA 02115-5000, USA} \\
%\footnote{ $\dagger$ : Permanent address}
\vskip 1.0 true cm
\centerline{~ Abstract}
\medskip
We discuss CP violation in supersymmetric theories and show that CP 
phenomena can act as  a probe of their origins, i.e.,  compactification 
and spontaneous supersymmetry breaking. CP violation 
as a probe of the flavor structure of supersymmetric theories is also 
discussed. A brief overview is given of several low energy phenomena 
where CP phases can produce new effects. These include important CP
effects in processes involving sparticles and CP mixing effects in the 
neutral Higgs boson system. We also discuss
the possibility of violations of scaling in the electric dipole moments
(EDMs) due to the presence of nonuniversalities and show that with 
inclusion of nonuniversalities the muon EDM could be up to
1-2 orders of magnitude larger  than implied by scaling and 
within reach of the next generation of experiments. Thus the  
EDMs are an important probe of the flavor structure of supersymmetric 
theories. 
\end{titlepage}

\section{Introduction}
In this talk we give a brief overview of recent developments on 
CP violation in supersymmetric theories.  
We begin with a discussion of the current experimental status of CP 
violation in the Standard Model\cite{ckm}.
 The electro-weak sector of the Standard 
 Model has one CP violating phase in the CKM matrix. An important constraint
on the Cabbibo-Kobayashi-Maskawa (CKM) matrix is that of unitarity and one such constraint is 
$ V_{ud}V^*_{ub} + V_{cd}V^*_{cb} + V_{td} V^*_{tb}=0$. 
 This  constraint can be represented 
by a unitarity triangle whose angles $\alpha, \beta, \gamma$  are 
defined by the relations
$ \alpha =arg(-{V_{td}V^*_{tb}}/{V_{ud}V^*_{ub}})$,
$\beta =arg(-{V_{cd}V^*_{cb}}/{V_{td}V^*_{tb}})$, and
$\gamma =arg(-{V_{ud}V^*_{ub}}/{V_{cd}V^*_{cb}})$
Currently 
there are four pieces of experimental evidence  for 
the existence of CP violation in nature.
Two of these come from the kaon system in the form of  $\epsilon$
and $\epsilon'/\epsilon$. A third piece of evidence appeared last year
from a direct measurement of $\sin (2\beta)$ in B meson decays
$B^0_d(\overline {B^0_d})\rightarrow J/\Psi K_s$ 
which gave $\sin 2\beta = (0.75\pm 0.10)$ (BaBar) and 
 $\sin 2\beta = (0.99\pm 0.15)$ (Belle). 
The fourth piece of evidence for CP violation is indirect
as it comes from the existence of baryon asymmetry in the universe
so that $n_B/n_{\gamma} =(1.5-6.3)\times 10^{-10}$.
Now the first three pieces of evidence appear to be consistent with the 
CP violation given by the Standard Model while the fourth one points 
to a new source of CP violation above and beyond the one from the
 standard model. CP violation also leads to electric dipole moments 
 of elementary particles. However, in the lepton sector of the 
 standard model EDMs arise at the multi 
loop level\cite{hoogeveen} and are typically more than ten orders of 
magnitude smaller than the current experimental limits\cite{commins,pdg} 
and thus for all practical purposes too small to be experimentally 
observed in the foreseeable future.
 This means that the observation of
a lepton EDM would be a clear indication of new physics beyond the 
standard model. In addition to the CP violation arising from the 
CKM matrix in the electro-weak sector of the Standard Model, 
one also has an additional CP violation given by the strong interaction 
of the theory, in the form of a term 
$\theta_G\frac{\alpha_s}{8\pi}G\tilde G$ which can give a huge EDM to
the neutron unless $\theta$ is fine tuned to be very small.

There is a large amount of literature now on how to suppress the
strong CP violation effects. Some of the early ideas discussed in this
connection consist of using axions, a massless up quark or a
symmetry argument to suppress CP violating effects\cite{barrnelson}.  
There is also considerable recent literature on this subject.
which we discuss briefly. The analysis of Ref.\cite{ma} invokes the
idea of a gluino-axino and uses a Peccei-Quinn mechanism for the 
gluino rather than the quarks. Thus it is  proposed that the axion 
couple to the gluino rather than the quark and that supersymmetry 
 breaking have the same origin as the axion. One finds  then that the 
$\theta_{QCD}$ is canceled exactly by the minimization of the
dynamical gluino phase and thus the calculation of the EDMs in MSSM
becomes unambiguous. The analysis of Ref.\cite{bdm1} is based on 
Left-Right symmeteric models. In these models the strong CP parameter 
$\bar\theta$ is zero  at the tree level, due to parity (P), 
but is induced due to P -violating effects
below the unification scale. It is estimated that 
$\bar \theta\leq 10^{-16}$ for models with universal scalar masses.
In more general SUSY breaking scenario, one finds 
$\bar\theta \sim (10^{-8}-10^{-10})$ close to experimental observation.
In the analysis of Ref.\cite{schmaltz} a solution to the strong CP problem 
using supersymmetry\cite{schmaltz} is proposed. Thus the work of 
 Ref:\cite{schmaltz} envisions a solution to the 
strong CP problem based on supersymmetric non-renormalization 
theorem. In this scenario CP is broken spontaneously and its breaking 
is communicated to the MSSM by radiative corrections. The strong CP phase is 
protected by a SUSY non-renormalization theorem and remains
exactly zero while the loops can generate a large CKM phase from
wave function renormalization. Finally in the analysis of Ref.\cite{ibanez}
a solution based on gauging away the strong CP problem is proposed.
Thus the work of Ref.\cite{ibanez} proposes a  solution that involves 
the existence of an unbroken gauged $U(1)_X$ symmetry whose gauge boson 
gets a Stuckelberg mass term by combining with a pseudoscalar field 
$\eta (x)$ which has a axion like coupling to $G\tilde G$. 
Thus the $\theta$ parameter can be gauged away by a $U(1)_X$ 
transformation. This leads to mixed gauge anomalies which are canceled
by the addition of an appropriate Wess-Zumino term.

\section{Large Phases in SUSY, String, and Brane Models }
Assuming that the strong CP problem is solved we still have
a SUSY CP problem.
Thus in the minimal supergravity model\cite{msugra,sugra} one has two 
arbitrary phases which are theoretically unrestricted and can get large. 
Similarly in a broad class of string and brane models one finds
that soft susy breaking generates CP violating phases of sizes which
are typically $O(1)$. Thus one finds that even in the absence of the
strong CP problem one has a SUSY CP problem in that in most models
based on SUSY, string and branes the phases of the soft parameters
are large and this is  problematic for the satisfaction of the 
EDM constraints. Many avenues have been explored on how to overcome 
this problem. Some possible solutions to this problem that have been
 discussed are the following:
(i) Arrange the satisfaction of the EDMs by fine tuning the phases to
be small\cite{ellis}.
(ii) Assume the phases are large but suppress their effects on the EDMs
of the quarks and the leptons by making masses of the sparticles
heavy\cite{na}.
Effectively this requires that the sparticle masses lie in the range
of several TeV. While this is a valid solution it appears  
contrary to the concept of  naturalness\cite{ccn}.
 (iii) Arrange so that the phases in the first 
two generations and the flavor blind phases vanish and the only phases
present are in the third generation\cite{chang}.
(iv)Internal cancellations\cite{incancel,rparity}:
Here one looks for regions of the parameter space where, one has
cancellations between the various contributions to the EDMs of the
leptons and of the quarks. If this mechanism holds then
one will expect to observe EDMs by a factor of 10 improvement 
in experiment. In addition to the above there are several variants of the
above ideas available in the literature\cite{dimo,bagger}. 
The EDM of the mercury $^{199}H_g$ is known to a great degree of
accuracy\cite{atomic} and one may 
wish to impose this constraint on the SUSY phases.
 However, atomic EDMs have several 
uncertainties which are not fully understood. These include uncertainties
associated with particle physics effects (such as the uncertainties 
associated with the strange quark content of the nucleons\cite{cn}), 
uncertainties associated with the nuclear physics models at
low energy, and 
uncertainties associated with atomic physics effects. In addition to the
above estimates made in computing the atomic EDMs have ignored the
effects of the CP violating dimension six operator\cite{Khatsymovsky}.
While the effect of the CP violating dimension six operator may be
small in some regions of the parameter space there is no reason to
believe it would be small in all allowed regions of the MSSM 
parameter space.

An interesting issue concerns the origin of CP violation. 
The two possibilities that present themselves are (1) compactification and 
 (2) spontaneous  symmetry breaking. Thus in the first case 
 while the string or M theory in its uncompactified form is CP symmetric 
 CP violation can appear after compactification. In such a 
 circumstance the Yukawa couplings are the ones likely to get
 complex phases\cite{cpstrings} which are eventually translated in
 terms of CP violation in the CKM matrix. Thus one can view the CP 
 violation in the CKM matrix as directly originating from the
 string compactification. Regarding the second possibility 
CP violation can arise when the spontaneous breaking of 
supersymmetry occurs. In supergravity and string 
models\cite{msugra,sugra,kj} this breaking arises also at the string/Planck scale. 
 Here one is led a priori to the presence of many more phases. 
(In gauge mediated breaking, or in M theory/brane  models the
  scale where soft CP phases appear could be in the 10 TeV region.)
There is yet another possibility for the origin of CP violation 
in the context of spontaneous breaking, and this is via 
spontaneous symmetry breaking in the electro-weak sector.
Now while the spontaneous CP violation does not occur in the Higgs sector of
MSSM it can occur in extensions of MSSM. Thus, for example, 
a recent analysis with an extensions of MSSM
 with the addition of two Higgs singlets 
 exhibits such a spontaneous breaking\cite{ham1}.
There are now the following possibilities: (a)
The SUSY contributions to K and B physics turn out to be small:
In this case one has a rather clean demarcation, that is 
the CP violations in K and B physics are probe of string compactification,
and baryogenesis and other CP phenomena that may be seen in sparticle
decays etc become a probe of spontaneous supersymmetry breaking.
(b) The second possibility is that the SUSY contributions to the
K and B systems are significant. In this case one has a more 
involved picture in that both $\delta_{CKM}$ and the SUSY phases 
contribute here and thus one would need to disentangle the SUSY CP 
effects from the effects of  $\delta_{CKM}$ to get any 
meaningful constraints on either $\delta_{CKM}$ or on the SUSY phases. 
On a more theoretical level one might ask if in MSSM there
exists a connection at some level between the CP violation in the
CKM matrix and the CP violation that arises in the soft parameters.
A priori there does not appear to be a connection as the origins of
these types of CP violations are very different, since one arises from 
string compactification and the other from spontaneous supersymmetry 
breaking. This is mostly true except that the soft trilinear
couplings $A_{\alpha\beta\gamma}$ have a dependence on the 
Yukawa couplings $Y_{\alpha\beta\gamma}$ via the 
a term of the form 
$A_{\alpha\beta\gamma}\sim F^i\partial_i Y_{\alpha\beta\gamma}$
where $i$ refer to the moduli fields. Thus in this case 
the CP violating phases in the  Yukawa couplings  will 
filter into the trilinear couplings $A_{\alpha\beta\gamma}$. 
However, the exact nature of compactification in string theory 
that takes us from 10 space time dimensions to 4 dimensions in 
string theory is not known. Similarly, the spontaneous breaking
of supersymmetry in string theory is not fully understood.
Thus in view of the above one cannot put strict theoretical 
restrictions on the size of CP violation either on $\delta_{CKM}$
that arises via string compactification or on the soft SUSY 
parameters that arise from spontaneous supersymmetry breaking.

\section{CP phases and low energy phenomena}
In the absence of any reliable string models that might in a 
natural fashion determine the CP phases to either vanish or be
small, one has to assume that these phases will in general be
sizable and that the resolution of the EDM problem comes about by one
of the methods discussed in the previous section.  In this 
circumstance when the phases are sizable a whole arrary of
 low energy phenomena will be 
affected. The list of such phenomena is indeed large and 
encompasses essentially all of current low energy phenomenology.
We list below a few of these. They include 
 sparticle masses and decay branching ratios and 
cross-sections\cite{mrenna,moretti,zerwas,bartl},
 Higgs boson decays, neutralino relic density and detection rates 
in dark matter detectors\cite{cin}, 
 g-2\cite{ing}, CP even -CP odd mixing in neutral 
Higgs system\cite{pilaftsis,carena,inhiggs1,inhiggs2}, 
 FCNC ~$b\rightarrow s+\gamma$\cite{cpbsgamma}, 
 trileptonic signal\cite{trilep},
 $\epsilon'/\epsilon$\cite{masiero1}, 
 CP effects on $e^+e^-$ and $\mu^+\mu^-$ collider 
phenomenology\cite{barger,AA,baek1,baek2},
CP effects on $b\bar b$ system\cite{voloshin}, 
 baryogenesis\cite{baryogenesis},  
 proton decay\cite{cpproton} and 
 $B^0_{s,d}\rightarrow \mu^+\mu^-$\cite{inbmumu}. 
Of the above we  will discuss the CP effects in the 
neutral Higgs in some detail in Sec.4 and the effects of 
SUSY CP phases on $g_{\mu}-2$ and on the muon EDM in Sec.5. 
Here we discuss briefly some general aspects of the phases.
One question one might ask is how one may determine the phases.
This is a more difficult task that might appear at the 
surface. The reason for this is that supersymmetric phenomena
in general depend typically on a combination of phases and
thus deciphering the phases would require a simultaneous 
determination of many observables. To see the complexity of 
this problem we exhibit the phase dependence of a few 
experimentally observable quantities. To streamline the 
discussion we begin by defining the soft parameters that enter
the low energy theory that will be the focus of our discussion.
The low energy theory in general consists of the following soft
parameters\cite{sugra}: sfermion masses $m_{\tilde f_L}, m_{\tilde f_R}$,
the U(1), SU(2), and SU(3) gaugino masses $m_i=|m_i|e^{i\xi_i}$ 
(i=1,2,3), the trilinear soft parameters $A_{f}$= $|A_{f}|e^{i\alpha_{A_{f}}}$,
and the Higgs mixing paramrter $\mu  =|\mu|e^{i\theta_{\mu}}$. 
Let us now consider the chargino and neutralino masses. The
chargino masses depend on the single combination $\xi_2+\theta_{\mu}$
while the neutralino masses depend on the phases $\xi_1, \xi_2, \mu$.
 The FCFC decay 
 $b\rightarrow s+\gamma$ depends on four combinations
$\xi_1+\theta_{\mu}$,$\xi_2+\theta_{\mu}$,$\xi_3+\theta_{\mu}$
and $\alpha_{A_t}+ \theta_{\mu}$.
 Similary, at least four combinations of phases appear in the 
decay of charginos $\tilde W\rightarrow q_1\bar q_2 +\chi_1$ which
may be taken to be  $\xi_1+\theta_{\mu}$,$\xi_2+\theta_{\mu}$,
$\alpha_{q_1}+ \theta_{\mu}$, and $\alpha_{q_2}+ \theta_{\mu}$.
Clearly then measurement of several quantities will be necessary in 
order to determine the phases. A relevant question then is the 
accuracy with which such phases can be determined. 
The linear collider is an appropriate instrument for
the accurate determination of the phases. 
Thus, for example, accurate measurements of the chargino and neutralino
 masses and their pair production cross sections at linear 
colliders can allow a determination of the CP violating phases to 
a good accuracy. With the design parameters of the linear colliders
a determination of some phases to an accuracy of one tenth of
a radian is possible\cite{barger}.  

\section{Effects of CP violation in the neutral Higgs sector}
At the tree level the Higgs mass matrix  in MSSM is block diagonal between 
the CP even and the CP odd odd states and here one gets two CP even 
states and one CP odd state. The situation changes when one 
includes the effects of loop corrections to the effective 
potential\cite{anloop}. 
In the presence of CP violating phases the loop corrections 
mix the CP even and the CP odd  Higgs sectors\cite{pilaftsis}
and the neutral Higgs mass matrix no longer factorizes into  
CP even and CP odd sectors. Thus in MSSM one has two Higgs 
doublets $(H_1^0, H_1^-)$ and $(H_2^+, H_2^0)$  where 
$H_2$ gives mass to the up quarks and $H_1$ gives mass to the
down quarks and the leptons. 
We denote the real and imaginary 
parts of the $H_1^0$ and $H_2^0$  by $\phi_1, \psi_1$,
and $\phi_2$ and $\psi_2$. After spontaneous breaking we go to the
basis $\phi_1, \phi_2, \psi_{1D}, \psi_{2D}$ defined by
$\psi_{1D}=\sin\beta \psi_1+ \cos\beta \psi_2$,
$\psi_{2D}=-\cos\beta \psi_1+\sin\beta \psi_2$. 
In this basis $\psi_{2D}$ decouples and one is left with 
a $3\times 3$ matrix which mixes the 2 CP even and one CP odd states.
It was first shown in Ref.\cite{pilaftsis} that the effect of top-stop 
exchange can generate a significant mixing between the CP even
and the CP odd states. In Ref.\cite{inhiggs1} it was extended to include
the chargino, $W$  and charged Higgs boson exchange and it was found this
exchange could  be very significant for large $\tan\beta$\cite{inhiggs1}.
it turns out that that a 
 similar computation to include the CP violation effects from
the neutralino, $Z$ and neutral Higgs  exchanges is far more difficult. 
The reason for this
is that while  the computation of the top-stop and charged Higgs, $W$ and
chargino exchange calculation involves diagonalization of only  2x2 
 mass matrices, the computation of the
neutralino exchange contribution involves the 
diagonalization of a 4x4 neutralino mass matrix. 
To deal with this problem one needs to use  some special techniques.
A full analysis of the neutralino exchange corrections with inclusion
of CP effects is given in Ref.\cite{inhiggs2}. Again one finds that
the effects of neutralino exchange contributions for large $\tan\beta$
can be quite significant comparable to the contributions from the
top-stop exchange contributions. 
There are a variety of signals associated with the effects 
of CP violation in the Higgs sector. One interesting signal 
associated with 
the CP even -CP odd mixing in the neutral Higgs sector can be 
seen directly in $e^+e^-$ colliders, e.g., three peaks in the process 
$e^+e^-, q\bar q\rightarrow Z^*\rightarrow Z+H_i$,
and  modified rates of $h\rightarrow b\bar b, 
c\bar c$ etc\cite{pilaftsis}.
Further, a very interesting observation in made in Ref.\cite{inhiggs3}
 is that if a mixing effect is observed experimentally
then among the three possibilities, i.e., the fine tuning,
the heavy sparticle spectrum, and the cancellation mechanism, 
it is only the cancellation mechanism that can survive under the 
naturalness constraint\cite{ccn}.
Another possible way to detect the existence of CP violation is in 
polarization asymmetries in the Higgs decays 
which have been analyzed recently\cite{drees}.
Specifically in this case one analyzes the spin correlated decays 
of the Higgs into neutralinos
and charginos and one can express the decays in terms of the 
longitudinal and the transverse polarization of the final charginos and
neutralinos. These spin correlated decays show a strong 
dependence on the CP violating phases. Thus an observation of 
the spin correlated decays can shed a considerable light on the
presence of supersymmetric CP violation in the Higgs sector.

\section{CP violation and flavor structure}
The contribution of the supersymmetric CP violation to the 
K and B systems can act as a probe of the flavor structure of 
supersymmetric models. Specifically one can envision three scenarios:
(a) Negligible or small contribution from the supersymmetric phases to
the K and B system: In this case essentially all of the CP violation in
K and B physics has standard model origin, i.e., arises from 
$\delta_{CKM}$ and K and B physics provides us with
no guide to the presence of a new flavor structure beyond what is 
present in the Yukawa couplings;
(b) Sizable contribution from SUSY phases to K and B physics:
If it turns out that there is a large contribution to the K and B physics
from supersymmetry, then in addition to large
supersymmetric CP phases, a new flavor structure beyond the Yukawas is
indicated\cite{masiero1,dine1}. For example, one needs a non negligible 
$(\delta_{12})_{LR}(d)= m^2_{LR}(d)/\tilde m_q^2 $
to get  a significant contribution to  $\epsilon'/\epsilon$
from supersymmetry; 
(c) The entire CP phenomena in K and B system arises from 
SUSY phases\cite{frere1,frere2}: As in (b) in this case  one  needs 
a new flavor structure in addition to large phases. However, 
there is no compelling reason for this extreme view point.
In the latter two scenarios, i.e., in (b) and (c) one  
can view CP violation as a probe of the flavor structure of 
the supersymmetric theory.

\section{CP violation, $g_{\mu}-2$ and the Muon EDM}
As mentioned already the muon anomalous moment is a very sensitive
function of CP phases and this effect has been analysed 
theoretically at considerable length in Ref.\cite{ing}.
It turns out that an experimental determination of the 
deviation of $g_{\mu}-2$ from the standard model also has
strong implications on constraining the CP phases. These
constraints were also analyzed in Ref.\cite{ing}.  
One very interesting result of the CP phases is that the usual
hierarchy between the chargino-sneutrino and the neutralino-smuon 
exchange contributions to 
$g_{\mu}-2$ no longer applies when phases exist. Thus in the
absence of CP phases one finds that the chargino-sneutrino 
exchange contribution to $g_{\mu}-2$ is typically much larger
than the neutralino-smuon exchange contribution. However, in
the presence of phases this does not necessarily hold. 
Thus while the chargino-sneutrino exchange contribution depends
only on the phase combination $\xi_2+\mu$,
 the neutralino-smuon  exchange contribution 
depends on the phases $\xi_1$, $\xi_2$, $\mu$ and $\alpha_{A_{\mu}}$.
Thus in this case depending on the numerical value
of the phases one finds that
the neutralino-smuon exchange contributions can become comparable to 
and even exceed the chargino-sneutrino exchange contribution.
The large sensitivity of $a_{\mu}^{SUSY}$ on the phases implies that 
a constraint on  $a_{\mu}^{SUSY}$ can be turned into
a constraint on the phases and this constraint turns out to be
quite significant if $a_{\mu}^{SUSY}\geq 10^{-10}$\cite{ing}. 
  Of course in the presence of CP phases there will be an EDM 
 associated with the muon and it is interesting to ask what the 
 size of this EDM is. This question is interesting in view of the
 fact that there is a recent proposal by BNL to probe 
 the muon EDM ($d_{\mu}$) with a sensitivity of  
$d_{\mu}\sim O(10^{-24})ecm$\cite{yanni}. This proposed limit
will be six orders of magnitude more sensitive that the current 
limit of $\sim 10^{-18}$ ecm,
Now for wide class of models one finds that a so called scaling 
holds so that the EDMs scale by the masses, i.e.,
 ${d_{\mu}}/{d_e}\simeq {m_{\mu}}/{m_e}$.
Since the electron EDM has an experimental  upper limit of 
 $d_e< 4.3\times 10^{-27}ecm$ scaling implies 
   $d_{\mu}<10^{-25}ecm$  which   
 lies below the sensitivity (i.e., $d_{\mu}<10^{-24}ecm$) that will be 
 accessible in the  proposed BNL experiment. 
 The question then is if there are any models which could
 produce a muon EDM sensitive to the proposed BNL experiment.
 Clearly this can happen only by a violation of scaling.
 Such scaling violations can occur in some Left-Right  symmetric
 models\cite{bdmedm}. More recently it was proposed that slepton
 nonuniversalities can generate such violations\cite{inmuedm,shadmi},
 These slepton non-universalities can easily arise from the 
 soft trilinear couplings, either in the magnitude or in phase.
 It is found that such nonuniversalities can easily generate
 a muon EDM in excess of  $10^{-24}ecm$ which is the sensitivity
 limit of the BNL experiment.

\section{Conclusions}
In this paper we have given a brief overview of the current 
status of CP violation in supersymmetric theories. 
We argued that CP violation arising from supersymmetric 
phases can influence a wide range of low energy phenomena.
These include  masses and  decays of sparticles, 
the neutral Higgs system, flavor changing processes such
as $b\rightarrow s+\gamma$, the trileptonic signal, 
the process $B^0_{s,d}\rightarrow  \mu^+\mu^-$, dark matter,
proton decay and many other processes. Clearly, the
CP phases then have an important impact on SUSY and 
Higgs searches. On a theoretical level, the origin of 
CP violation can be traced back to string compactification and
to the spontaneous breaking of supersymmetry. The CP violation 
associated with Yukawa couplings that arise as a consequence of 
string compactification largely affects the CKM matrix, while
phenomena such as the EDMs, low energy SUSY phenomenology and the 
baryon asymmetry of the universe are largely governed by the
CP phases that are associated with the spontaneous breaking of 
supersymmetry. Assuming  K and B physics is largely controlled by
the CP violation in the CKM matrix, such physics gives us a 
glimpse into the nature  of string compactification, while 
CP violation in low energy SUSY and Higgs  phenomena which we
expect to observe in collider experiments will reveal the nature 
of CP violation associated with soft parameters, and such physics
will give us a glimpse into the nature of spontaneous breaking of
supersymmetry.

\section{Acknowledgments}
This research was supported in part by NSF grant PHY-0139967.

\end{document}